\documentclass[aps,twocolumn,amsmath,amssymb,showpacs,prl,floatfix]{revtex4}
\usepackage{amsmath}
\usepackage{graphicx}
\usepackage{epsfig}
\usepackage{psfrag}
\newlength{\upit}\upit=0.1truein

\begin{document}
\title{ Non-Fermi liquid behavior with and without quantum criticality in Ce$_{1-x}$Yb$_x$CoIn$_5$}

\author{T. Hu,$^{1,*}$  Y. P.  Singh,$^{1,*}$  L. Shu,$^2$ M. Janoschek,$^2$ M. Dzero,$^1$  M. B. Maple,$^2$ and C. C. Almasan,$^1$}

\affiliation{$^1$Department of Physics, Kent State University, Kent, OH 44242, USA \\
$^2$Department of Physics, University of California at San Diego, La Jolla, CA 92903, USA}
\date{\today}\vspace{0.5cm}

\maketitle

{\bf One of the greatest challenges to Landau's Fermi liquid theory - the standard theory of metals - is presented by complex materials with strong electronic correlations. In these materials, non-Fermi liquid transport and thermodynamic properties are often explained by the presence of a continuous quantum phase transition which happens at a quantum critical point (QCP). A QCP can be revealed by applying pressure, magnetic field, or changing the chemical composition. In the heavy-fermion compound CeCoIn$_5$, the QCP is assumed to play a decisive role in defining the microscopic structure of both normal and superconducting states.  However, the question of whether QCP must be present in the material's phase diagram to induce non-Fermi liquid behavior and trigger superconductivity remains open. Here we show that the full suppression of the field-induced QCP in CeCoIn$_5$ by doping with Yb has surprisingly little impact on both unconventional superconductivity and non-Fermi liquid behavior. 
This implies that  the non-Fermi liquid metallic behavior could be a new state of matter in its own right rather then a consequence of the underlying quantum phase transition.}

The heavy-fermion material CeCoIn$_5$ is a prototypical system in which strong interactions between conduction and predominantly localized $f$-electrons give rise to a number of remarkable physical phenomena \cite{PiersReview,SarraoReview}. Unconventional superconductivity emerges in CeCoIn$_5$ out of a metallic state with non-Fermi-liquid (NFL) properties: linear temperature dependence of resistivity below 20 K, logarithmic temperature dependence of the Sommerfeld coefficient, and divergence of low temperature magnetic susceptibility \cite{LinearT,HeatCapacity,SpecificHeat, logT}.  These anomalies disappear beyond a critical value of the magnetic field and the system recovers its Fermi liquid  properties. The crossover from non-Fermi liquid to Fermi liquid behavior is thought to be governed by a quantum critical point (QCP), which separates paramagnetic and antiferromagnetic (AFM) phases and is located in the superconducting phase \cite {Crossover,Singh}. Neutron scattering studies \cite{GegenScience2007} and more recent measurements of the vortex-core dissipation through current-voltage characteristics \cite{Our} provide direct evidence for an antiferromagnetic QCP in CeCoIn$_5$ that could be accessed by tuning the system via magnetic field or pressure.

Nevertheless, a growing number of $f$-electron systems do not conform with this QCP scenario; for example, the NFL behavior in some systems occurs in the absence of an obvious QCP \cite {betaYbAlB4, MapleReview}. An intriguing candidate is Yb-doped CeCoIn$_5$ that exhibits an unconventional $T-x$ phase diagram without an apparent QCP, while the onset of coherence in the Kondo lattice and the superconducting transition temperature $T_c$ are only weakly dependent on Yb concentration and prevail for doping up to $x = 0.65$. \cite{Yb-valence}. Yet, the presence of a QCP in the parent CeCoIn$_5$ compound and the logarithmic temperature dependence of normal state Sommerfeld coefficient in lightly doped  Ce$_{1-x}$Yb$_x$CoIn$_5$ crystals \cite{Wilson2} show that this system is in the vicinity to a QCP. 
Therefore, it is important to elucidate the nature of the NFL behavior and unconventional superconductivity in such a system, to search for possible QCPs,  and to determine the degree to which quantum criticality and superconductivity are coupled to each other. 

\begin{figure}[!]
\includegraphics[width=3.2in,angle=0]{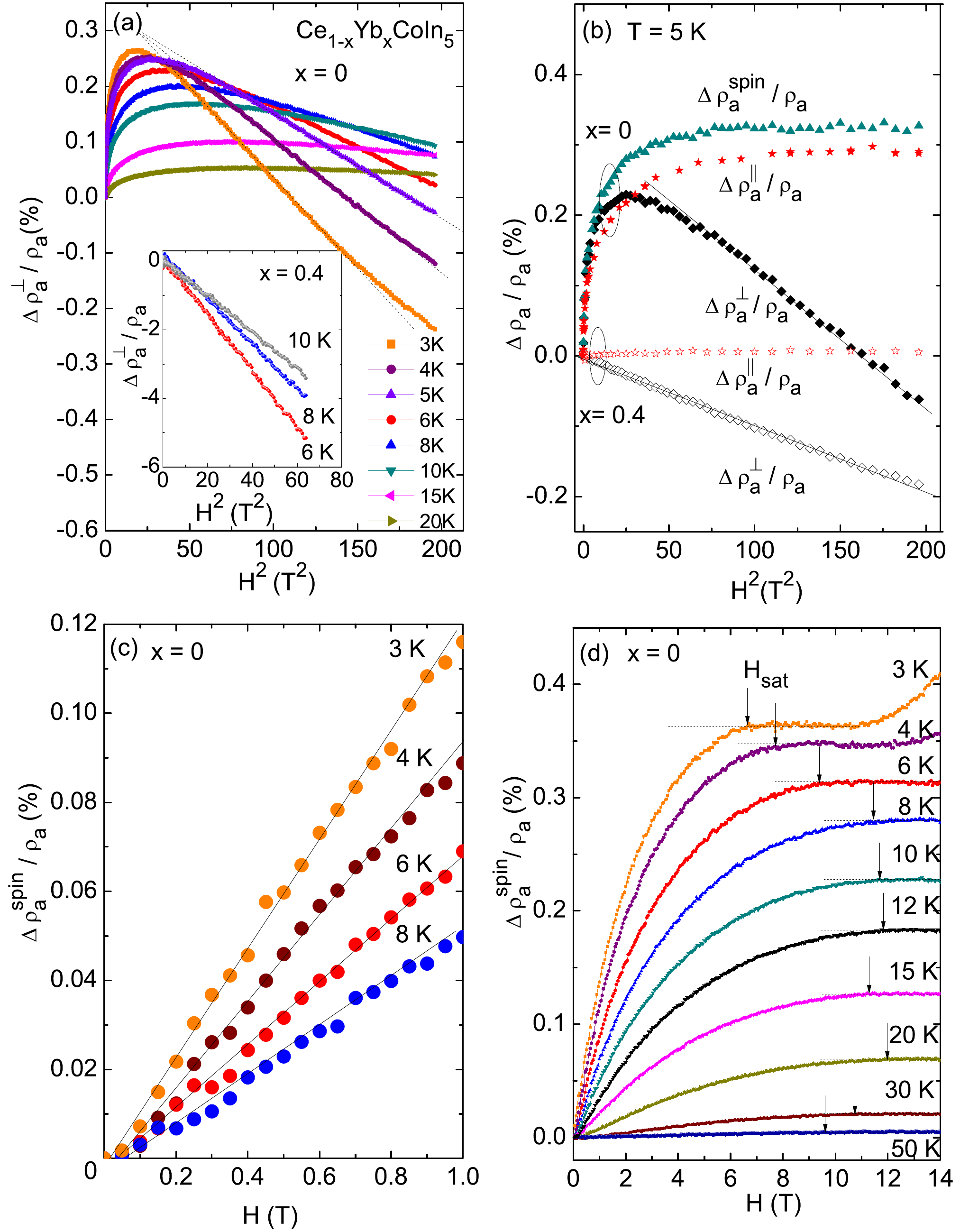}
\caption{(Color online) (a) Magnetic field $H$ scans of transverse magnetoresistivity $\Delta \rho_{a}^{\perp}/\rho_a$ measured at different temperatures on Ce$_{1-x}$Yb$_x$CoIn$_5$ for $x=0$ (main panel) and $x=0.4$ (inset). (b) In-plane transverse $\Delta \rho_{a}^{\perp}/\rho_a$  (black) and longitudinal $\Delta \rho_{a}^{\parallel}/\rho_a$ (red) magnetoresistivities (MR) of $x=0$ and $x=0.4$ samples vs $H^2$ measured at 5 K. The green data are denoted $\Delta \rho_{a}^{spin}/\rho_a$ and represent the component of MR obtained by subtracting the high field quadratic MR from the measured transverse MR. (c) $\Delta \rho_{a}^{spin}/\rho_a$ vs $H$ (for low $H$ values) for the $x=0$ sample measured at different temperatures. (d) $H$ dependence of the isotropic component $\Delta \rho_{a}^{spin}/\rho_a$ of transverse magnetoresistivity of CeCoIn$_5$ measured at different temperatures and up to 14 T.}
\label{Fig1}
\end{figure}
Here we focus on revealing a field-induced quantum critical point through normal-state magnetoresistivity measurements and finding its evolution with Yb-doping. Experimental details are given in the "Materials and Methods" Section. We first present the results of our study of the magnetic field $H$ and temperature dependence of the transverse  ($H  \perp ab$) in-plane  magnetoresistivity $\Delta \rho_{a}^{\perp}/\rho_a \equiv [\rho_a^{\perp}(H)- \rho_a(0)]/\rho_a(0)$, for $H \leq 14$ T and $3 \leq T \leq 70$ K. Figure 1(a) and its inset display the field dependence of $\Delta \rho_{a}^{\perp}/\rho_a$  measured at different temperatures for the $x = 0$ and $x = 0.4$ samples, respectively. We note that the data for these samples fall into two groups: (i) non-monotonic field dependence of magnetoresistivity (MR) [Fig. 1(a)] with quadratic MR at high fields, typical for $x\leq 0.20$ and (ii) negative and quadratic MR over the whole measured field range [inset to Fig. 1(a)], typical behavior for the high Yb ($0.25 \leq x \leq 0.65$).  We, therefore,  conclude that the MR for low Yb doping has two main contributions: one negative and quadratic in $H$, which we denote $\Delta \rho_{a}^{orb}/\rho_a$ [black symbols of Fig. 1(b) at high fields] and the other is {\it positive} denoted $\Delta \rho_{a}^{spin}/\rho_a$ [green symbols of Fig. 1(b)], with the latter one obtained by subtracting for all field values the negative quadratic MR from the measured MR. This latter contribution to MR is isotropic since the green data follow quite well the longitudinal magnetoresistivity $\Delta \rho_{a}^{\parallel}/\rho_a$ [red data of Fig. 1(b)]. Also, $\Delta \rho_{a}^{spin}/\rho_a$ is linear in $H$ at low fields [Fig. 1(c)] and saturates at high fields [Fig. 1(d)].  

\begin{figure}[!]
\includegraphics[width=3.2in,angle=0]{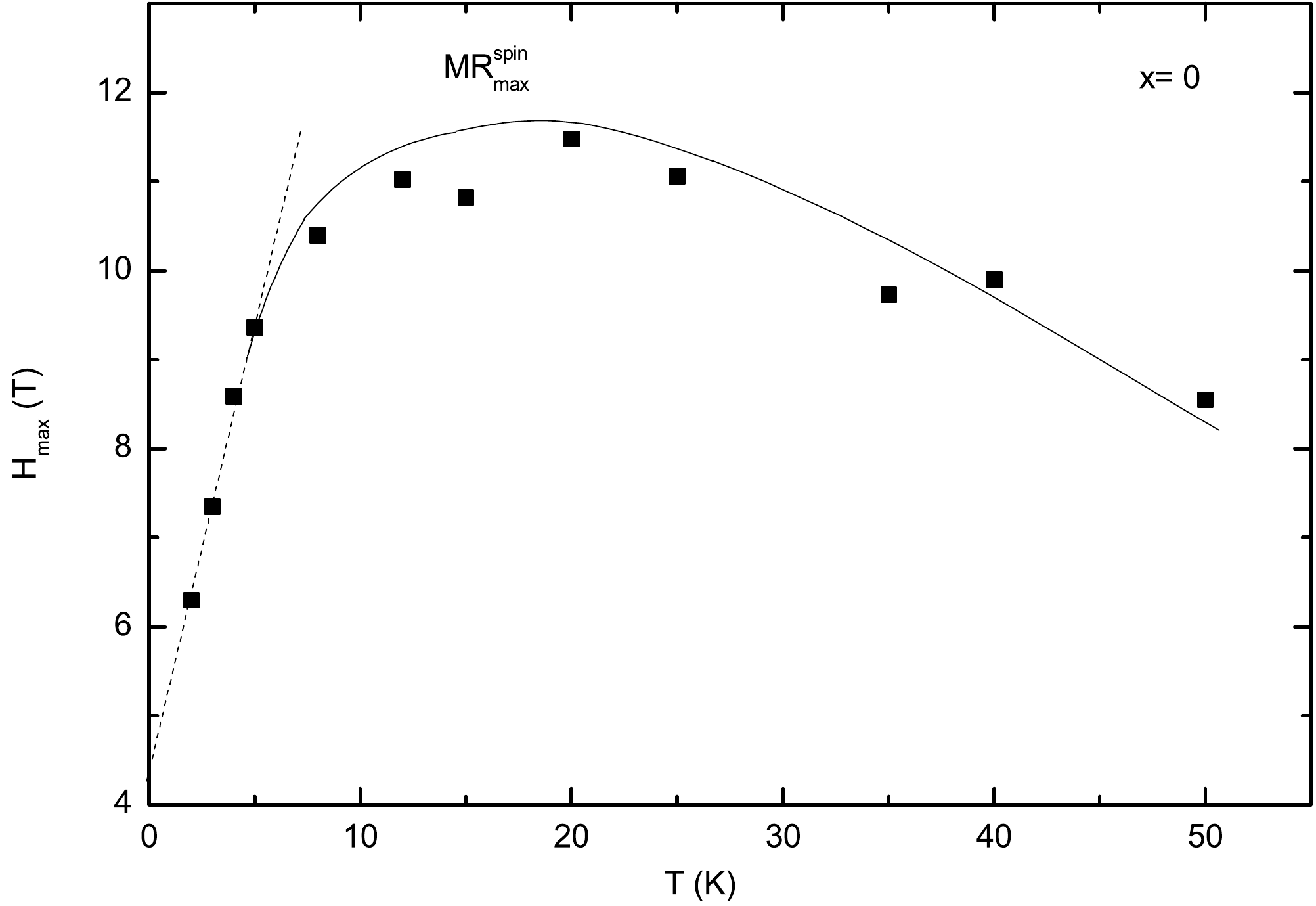}
\caption{Temperature dependence of the characteristic field $H_{max}$  corresponding to the maximum isotropic component of magnetoresistance $MR_{max}^{spin}$. The solid line is a guide to the eye, while the dotted line is a linear fit of the low-$T$ linear behavior of $H_{max}$for the $x=0$ sample.}
\label{Fig2}
\end{figure}
 
Positive MR of heavy-fermion materials at low fields marks the departure from the single-ion Kondo behavior and is determined by the formation of the coherent Kondo lattice state in systems in or close to their Fermi-liquid ground state 
\cite{CeCu6exp,UPt3,MagnetoExp,SteglichMagneto,Aronson1989,Theory1}.  
The maximum in the MR of a Kondo-lattice Fermi liquid at a certain field value is a result of the competition between a $T$-independent residual resistivity contribution that increases with increasing $H$, and a $T$-dependent term that decreases with increasing $H$ \cite{Theory1, Chen}. Thus, in conventional Kondo lattice systems, the peak in the field-dependent MR moves toward lower $H$ with increasing $T$ since  a lower field is required to break the Kondo lattice coherence.  

To determine the nature of the positive magnetoresistance in Ce$_{1-x}$Yb$_x$CoIn$_5$ for $x \leq0.20$,  we extract the temperature dependence of the field $H_{max}$ at which $\Delta \rho_{a}^{spin}/\rho_a$ saturates at its maximum value $MR_{max}^{spin}$ [see Fig. 1(d)]. The data of Fig. 2 show that the temperature dependence of $H_{max}$ of the $x=0$ sample is non-monotonic, with a maximum around 20 K, and a linear behavior at low-$T$ values. This non-monotonic $T$ behavior of $H_{max}$ is representative for all the samples with $x \leq 0.20$. The positive MR measured at $T>20$ K [for which $H_{max}(T)$ decreases with increasing $T$] could reflect the presence of the coherent Kondo lattice state at low field values as discussed in the previous paragraph. In contrast, the behavior below 20 K  is opposite with the one discussed above for conventional Kondo lattice systems. The increase of $H_{max}$ with increasing $T$ at these lower temperatures had previously been  observed in CeCoIn$_5$ and it has been attributed to field quenching of the AFM spin fluctuations responsible for the NFL behavior \cite{MagnetoCeCoIn5}.  Therefore, we conclude that the positive MR measured at $T<20$ K in Ce$_{1-x}$Yb$_x$CoIn$_5$ with $x\leq 0.20$ reflects the dominant role played by the AFM quantum spin fluctuations.  

An important next goal is to identify the quantum critical field ($H_{QCP}$) associated with these quantum fluctuations. One option is to extrapolate the low temperature linear $H_{max}(T)$ behavior to $T=0$ K and identify this $H_{max}(0)$ with $H_{QCP}$. However, since there is a certain error associated with the determination of $H_{max}$, we adopted a more accurate procedure to unambiguously determine $H_{QCP}$ for different Yb doping. This procedure is discussed in detail in the "Materials and Methods" Section below. We show in Fig. 3(a) the values of $H_{QCP}$ as a function of Yb concentration. As expected, the value of $H_{QCP}$ of 4.1 T for CeCoIn$_5$ coincides  with the value of $H_{QCP}$ determined previously from both  resistivity measurements done in the normal state \cite{Conventional} and  $I-V$ characteristics measured in the mixed state \cite{Our}. Therefore, the measurement of $\Delta \rho_{a}^{\perp}/\rho_a$ along with the analysis used here represent an  excellent experimental technique to determine the field-induced QCP in the NFL regime. The $H_{QCP}=0$ points in Fig. 3(a) for $x>0.20$ correspond to the case when there is no positive MR, i.e., the maximum in MR shifts to zero field [see inset to Fig. 1(a)]. 

Figure 3(a) also shows the suppression of $T_c$ (from \cite{Yb-valence}) and $T_{coh}$ with doping for Ce$_{1-x}$Yb$_x$CoIn$_5$. These doping-dependent $H_{QCP}$ and temperature phase diagrams show that while superconductivity (SC) is robust and survives over the whole Yb doping range, the field-induced QCP is strongly suppressed with Yb doping and disappears for $x>0.20$. This implies that SC and quantum criticality are likely to be decoupled in this system, i.e., unconventional superconductivity is not triggered by spin fluctuations.

The experimental technique used to determine $H_{QCP}$ also permits the determination of the gyromagnetic factor $g$(see "Materials and Methods" Section below).  There is a  significant change in the value of the $g$ factor at QCP, from 2.2 just below $H_{QCP}$  to 1.3 just above $H_{QCP}$. This means that for $H<H_{QCP}$ the conduction electrons become weakly coupled to the local spins, so that the heavy-fermions would have a lower effective mass. However, for $H>H_{QCP}$ the antiferromagnetic fluctuations between the local moments are suppressed, so that the heavy-fermions recover their effective mass and this is reflected in the reduction in the value of the $g$-factor \cite{Coleman2001}.

\begin{figure}[!]
\includegraphics[width=3.2in,angle=0]{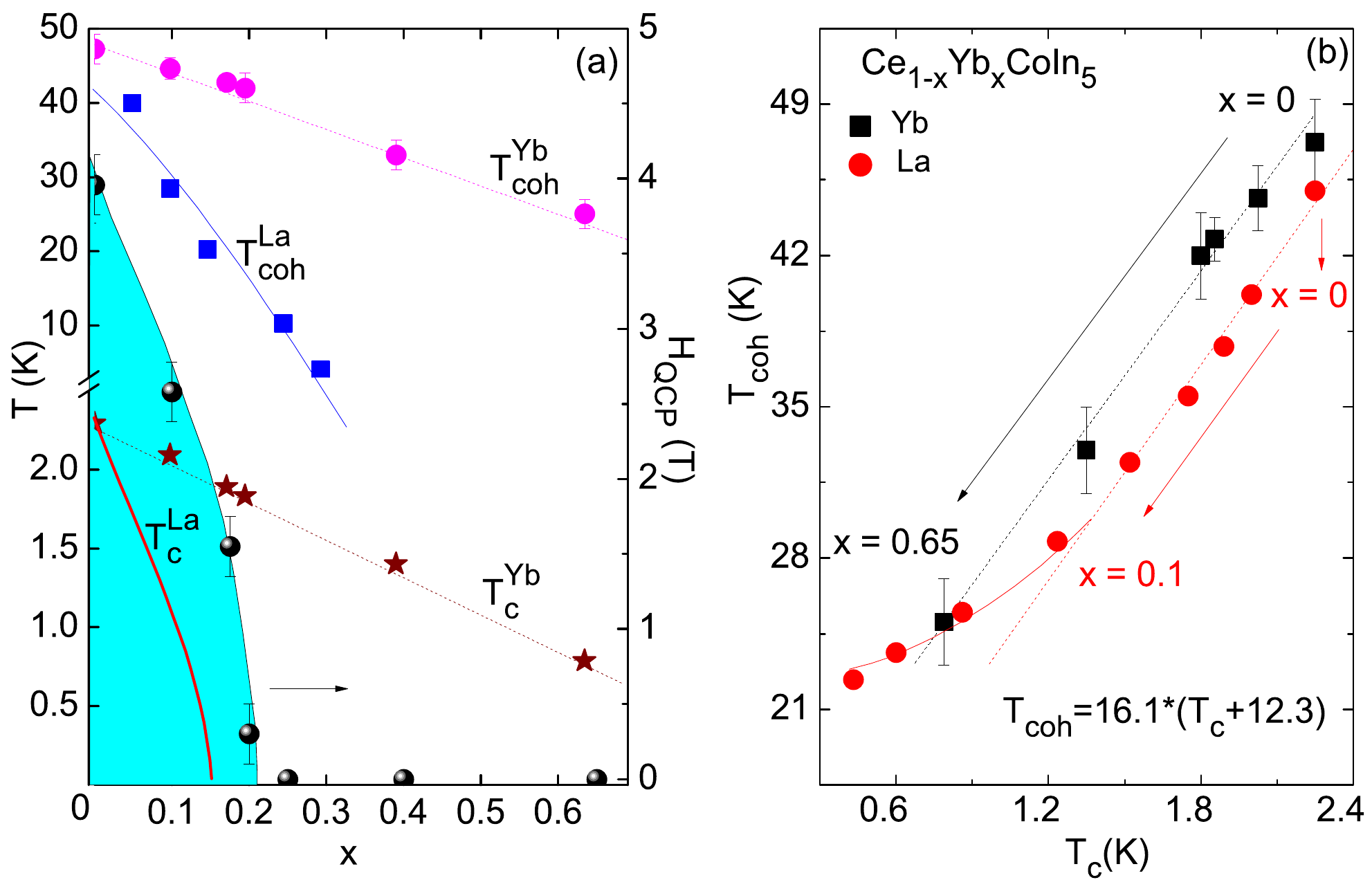}
\caption{(Color online) (a): Evolution of the field induced quantum critical point $H_{QCP}$, coherence temperature $T_{coh}$, superconducting critical temperature $T_c$ (from Ref.  \cite{Yb-valence}) of Ce$_{1-x}R_x$CoIn$_5$ ($R = $Yb, La) as a function of rear-earth  concentration $x$. (b): Plot of $T_{coh}$ vs. $T_{c}$ for Yb- and La-doped (from Ref. \cite{LinearT}) samples.}
\label{Fig3}
\end{figure}

Figure 3(b) shows that T$_{coh}$ (from this work) scales with $T_c$ (from Ref. \cite{Yb-valence}) over the whole Yb doping studied here ($x\leq 0.65$) (black squares) and La doping (red circles, data from Ref. \cite{LinearT}). We note that $T_{coh}$ vs $T_c$ follows a linear dependence for the whole Yb doping  with $x\leq 0.65$ and up to about 10\% La doping, with the two slopes having about the same value; this figure also shows the result of the linear fit for the Yb doping. Thus, it is intriguing that while the scaling of these two temperatures is in conformity to the other rare-earth substitutions on the Ce site \cite{JPNature}, the nature of the scaling sets the Yb substitution aside from the other rare-earth substitutions since, in the latter ones a non linear and quite fast suppression of both temperatures is observed, with a suppression of $T_c$ to zero at around $x=17.6$\% of rare-earth substitution (see Fig. 3(a), data from \cite{Yb-valence, LinearT}). 

A very interesting and puzzling behavior of the Ce$_{1-x}$Yb$_x$CoIn$_5$ system is that, even though QCP disappears for $x>0.20$, the system continues to display NFL behavior, as evidenced by the sublinear $T$-dependence of its resistivity \cite{Yb-valence}. This means that this non-Fermi liquid behavior at higher Yb doping could be a new state of matter in its own right rather then a consequence of the underlying quantum phase transition. We further investigated the origin of this NFL behavior by studying in more detail the $T$ dependence of the resistivity measured in different magnetic fields. The resistivity of all the Yb-doped single crystals studied follow remarkably well the expression $\rho_a^{\perp}(H) = \rho_0 + AT+B \sqrt T$ for temperatures up to about 15 K [see inset to Fig. 4(a) for the fitting results] and in fields up to 14 T; here $\rho_0$, $A$, and $B$ are doping- and field-dependent fitting parameters. We show in Figs. 4(a) and 4(b) the doping and magnetic field dependences, respectively, of the ratio of the $T$-linear contribution to the total $T$-dependent contribution of the resistivity. These results show that the linear in $T$ term in resistivity is present for low Yb doping ($x\leq 0.20$) in the quantum critical regime ($T\leq 20$ K). The percentage of the linear in $T$ term decreases with increasing $x$ and $H$ and disappears for $x>0.20$ and at fields at which the MR is only negative, where only the $\sqrt T$ dependence is present. 
The bottom inset to Fig. 4(a) shows the average Yb valence as a function of Yb doping \cite{Yb3}. It Is noteworthy that the average Yb valence decreases with increasing Yb and saturates to a value of about 2.3 for $x>0.20$ \cite{Yb3,Yb2.3}.  This result along with the data of Figs. 4(a) and 4(b) and the fact that both the superconducting transition and coherence temperatures remain weakly dependent on doping indicate that Yb atoms form a cooperative mixed-valence state that significantly reduces the pair-breaking effects, which could also play an important role in the origin of the NFL behavior at these higher Yb concentrations ($x>0.20$). This idea is supported by the observed linear dependence of $T_{coh}$ vs $T_c$  [Fig. 3(b)]. The "Materials and Methods" Section gives more discussion and interpretation of this scaling.

\begin{figure}[!]
\includegraphics[width=3.2in,angle=0]{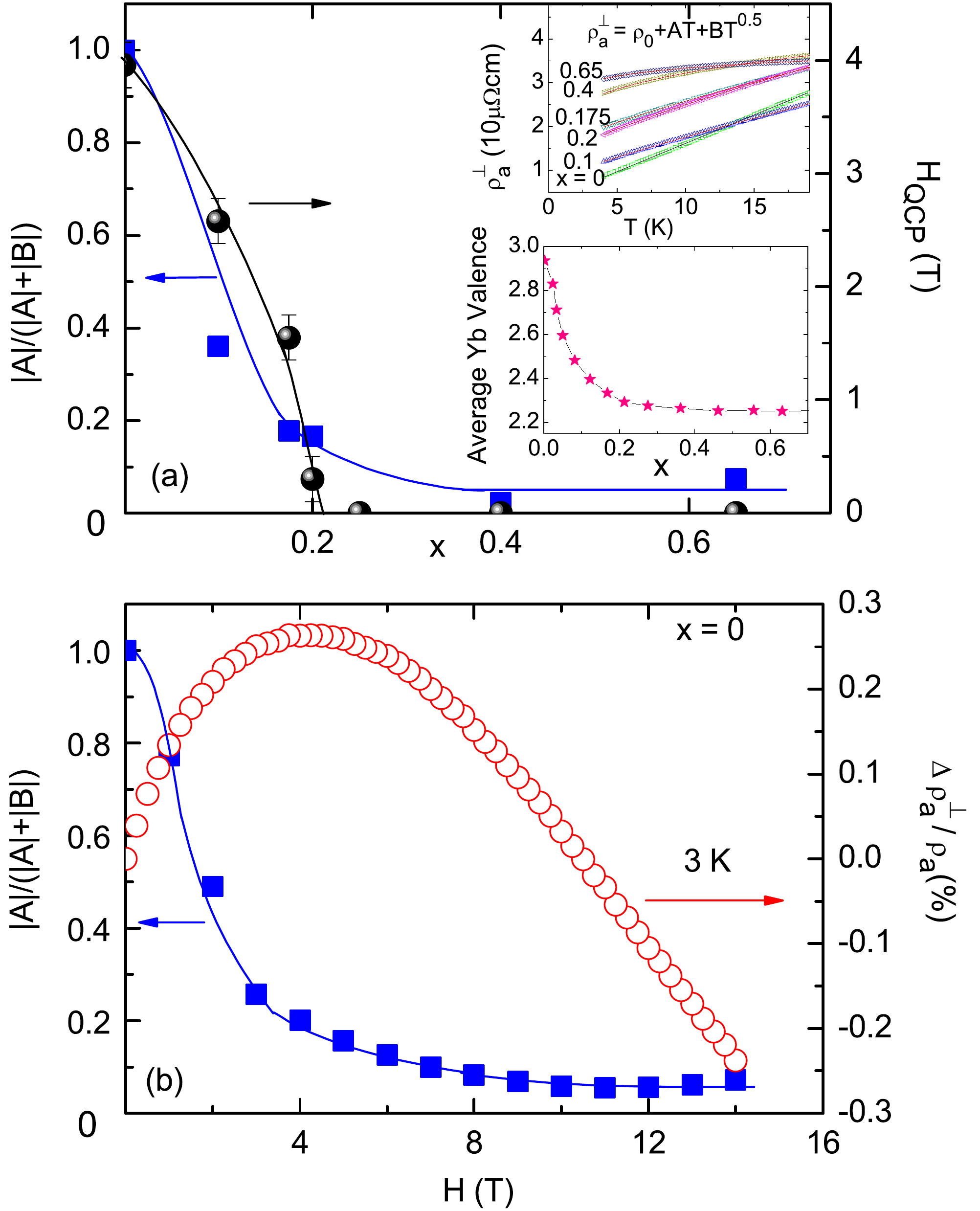}
\caption{(color online) (a) Doping dependence of linear in temperature T contribution devided by the total contribution to the resistivity obtained from fits (shown in the top inset) of the resistivity data by $\rho_a^{\perp}(H) = \rho_0 + AT+B \sqrt T$, along with the evolution of the field-induced QCP $H_{QCP}(x)$ (main panel). Bottom inset: Variation of average Yb valence with doping (data from Ref. \cite{Yb3}). (b) Field dependence of the linear in T contribution over the total contribution to the resistivity (left axis) and magnetoresistivity (right axis) measured at 3 K for pure CeCoIn$_5$.}
\label{FigS1}
\end{figure}

All of the above observations show that the field-induced QCP plays a supporting rather then competing role in the emergence of unconventional superconductivity. Moreover, we conclude that a highly unusual microscopic mechanism of unconventional superconductivity in Ce$_{1-x}$Yb$_x$CoIn$_5$ may be at play: on one hand, the normal state is fully reconstructed by the ytterbium substitution, while on the other hand, the superconducting critical temperature is reduced only by a factor of two at $x\simeq 65\%$, strongly suggesting that the superconducting pairing is spatially inhomogeneous and  involves local Ce $f$-moments. Finally, we propose a novel technique to probe the interplay between quantum criticality and superconductivity, which can be used to analyze a variety of strongly correlated electronic materials. 

\section{Materials and Methods}
In this Section we provide the supplementary discussion for the background material that will enable one to reproduce our results. First, we outline how the samples were prepared and describe the experimental methods used to determine the magnetoresistance and its dependence on magnetic field and temperature. Then, we show how describe the procedure we developed to determine the field-induced quantum critical point from magnetoresistivity data. Finally, we offer an interpretation of the scaling of the Kondo lattice coherence temperature and superconducting critical temperature.
\subsection{Experimental methods}
Single crystals of Ce$_{1-x}$Yb$_x$CoIn$_5$ ($0\leq x \leq 0.70$) were grown using an indium self - flux method \cite{Self_flux}. The crystal structure and composition were determined from X-ray powder diffraction (XRD) and energy dispersive X-ray (EDX) techniques. The single crystals have a typical size of $2.1 \times 1.0 \times 0.16$  mm$^3$, with the $c$ axis along the shortest dimension of the crystals. They were etched in concentrated HCl for several hours to remove the indium left on the surface during the growth process and were then rinsed thoroughly in ethanol. Four leads were attached to the single crystal, with $I \parallel a$. High quality crystals were chosen to perform in-plane transverse ($\Delta \rho_{a}^{\perp}/\rho_a$) and longitudinal ($\Delta \rho_{a}^{\parallel}/\rho_a$) magneto-resistivity (MR) measurements, with $H \perp ab$ and $H \parallel ab$, respectively, as a function of temperature ($T$) and applied magnetic field ($H$). In both situations, however, the field is perpendicular to the current, ${\vec H}\perp {\vec I}$, to ensure that Lorentz force remains the same.

\subsection{Determination of field-induced quantum critical point}
We used to following procedure to determine the quantum critical field $H_{QCP}$ for different Yb doping. We define the characteristic fields $H_{max}$ where $\Delta \rho_{a}^{spin}/\rho_a$ is $100\% $, $95\% $, $80\% $, $75\% $, etc. of the saturation value MR$_{max}^{spin}$.  We show the $T$ dependence of these characteristic fields  in Fig. 5(a) by the black, red, green, and blue solid circles, respectively. We then fit the linear low-$T$ behavior of these $H_{max}(T)$. Figure 5(b) is  a plot of the slopes $K$ (black solid circles), obtained from these linear fits of the different  curves with different percentages of MR$_{max}^{spin}$ [Fig. 5(a)], and the corresponding percentage of MR$_{max}^{spin}$ (red solid squares) vs the corresponding values of the intercept fields $H_{max}(0)$. 
We note the sharp increase of the slope $K$, at a certain $H_{max}(0)$ value. 

\subsection{g-factor and Kondo breakdown}
The origin of the sharp increase in the values of $K$ can be interpreted as follows. For a system in the quantum critical regime [i.e., low $H$ and $T$ data of Fig. 5(a)], the only energy scale is Boltzmann energy ${\cal E}_T=k_BT$. We compare this energy scale ${\cal E}_T$ to the quasiparticle Zeeman energy, i.e.,
\begin{equation}\label{HTQCP}
k_BT=g\mu_B(H-H_{QCP}).
\end{equation} 
From Eq. (\ref{HTQCP}) we see that the slope $K$ [Fig. 5(a)] must be inverse proportional to the gyromagnetic factor $g$. So, the sharp increase in $K$ is a result of the sharp decrease in $g$.  Previous studies \cite{gFactor} have shown that abrupt changes in the values of the gyromagnetic factor occur at the quantum critical point. Therefore, using this procedure we are able to unambiguously determine $H_{QCP}$ as the value of $H_{max}(0)$ at which there is the sharp change in the $g$ factor. Specifically, we find that $g$ decreases from 2.2 just below $H_{QCP}$  to 1.3 just above $H_{QCP}$.

The experimentally observed changes in the $g$-factor reflect the transformations that the electronic system undergoes under the 
change in external magnetic field. Given that $g\simeq 2.2$ just below $H_{QCP}$, numerically close to the value found for the free electrons in a metal, suggests that  the conduction electrons become decoupled from the local $f$-moments below the QCP. Thus, this observation can be interpreted using the phenomenological theory of "Kondo breakdown"  at $H_{QCP}$ \cite{KondoBreak}.  Within this theory, the changes in the $g$-factor are governed by the changes in the size of the Fermi surface: larger values of the $g$-factor correspond to small Fermi surface so that the conduction electrons are effectively decoupled from the localized $f$-states. In the opposite limit of smaller $g$-values, the Fermi surface is large, reflecting the strong coupling between the conduction and $f$-electrons. More importantly, the jump in the size of the Fermi surface at $H_{QCP}$ corresponds to the divergence of the quasiparticle's effective mass. 
\begin{figure}[!]
\includegraphics[width=3.2in,angle=0]{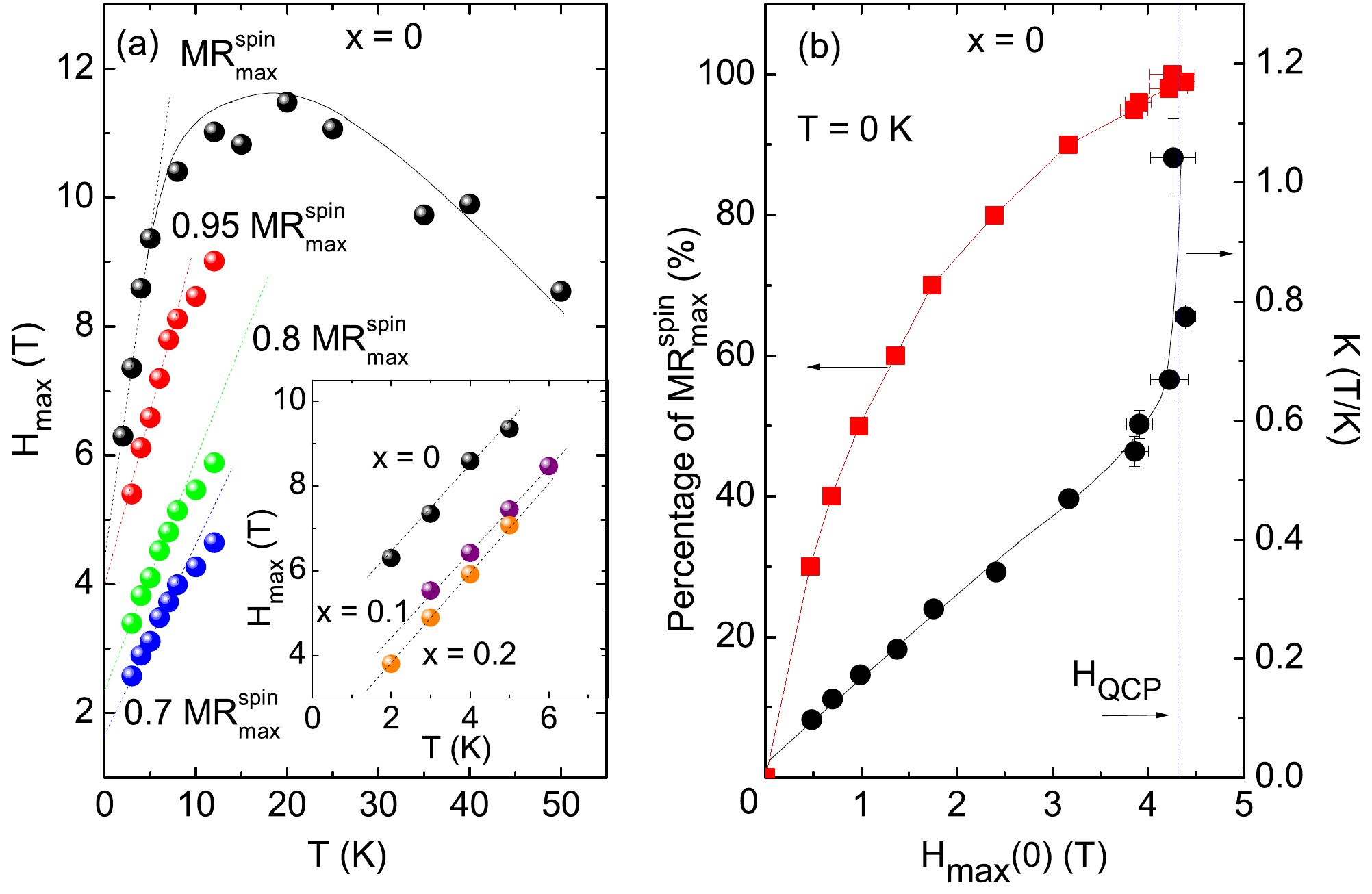}
\caption{(Color online) (a): Temperature dependence of the characteristic field $H_{max}$ determined for different percentages of the maximum isotropic component of magnetoresistance $MR_{max}^{spin}$ (main panel) and low-$T$ linear behavior of $H_{max}$ for $0 \leq x \leq 0.20$ samples (inset) (b): Slopes $K$ (black solid circles) and percentage of MR$_{max}^{spin}$ as a function of the corresponding values of the intercept fields $H_{max}(0)$.
}
\label{Fig1s}
\end{figure}
\section{Scaling of the coherence temperature}
In this Section, we discuss and offer an interpretation of our data for the concentration dependence of the Kondo lattice coherence temperature ($T_{coh}$) and superconducting critical temperature ($T_c$). 
For definiteness, we focus primarily on the concentration region $x\leq 0.5$, where $T_{coh}$ can be interpreted
as a heavy-fermion coherence temperature of the Ce $f$-moments. 

We start with the following observation: by combining our results on the concentration dependence of the coherence temperature and the critical temperature of the superconducting transition, we find that to a good accuracy both of them are suppressed at the same rate, i.e.,
\begin{equation}\label{ScalingYb}
T_{coh}^{Yb}(x)=\alpha_{Yb}T_c^{Yb}(x)+\beta_{Yb}, 
\end{equation}
where
\begin{equation}
\begin{split}
\alpha_{Yb}=16.1\pm0.2, \quad \beta_{Yb}= 4.09 T_{coh}(x=0).
\end{split}
\end{equation}
To get better insight into the physical meaning of this result, it is instructive to compare Eq. (\ref{ScalingYb}) with the same dependence obtained for 
La-substituted CeCoIn$_5$. We find that in this case $T_{coh}$ also scales with $T_c$ at small enough concentration of $La$ [Fig. 2(b) in the main text]:
\begin{equation}
T_{coh}^{La}(x)=\alpha_{La}T_c^{La}(x)+\beta_{La}.
\end{equation}
The slight offset between the two curves, Fig. 2(b), appears likely because of atomic size differences between Yb and La ions. 
What is truly surprising is that the rate it which both $T_{coh}$ and $T_c$ are suppressed turns out to be the same for both types of substitutions:
\begin{equation}\label{ratios}
\alpha_{Yb}\simeq\alpha_{La}.
\end{equation}
We interpret Eq. (\ref{ratios}) as an indication that the onset of the many-body coherence in the Kondo lattice and emergence of superconductivity
has the same physical origin: hybridization between conduction and localized $f$-electron states.  In particular, this suggests that Cooper pairing develops primarily on the "heavy" (i.e., large) Fermi surface. 

Note that, although the suppression rate of Kondo lattice coherence and superconductivity for both Yb- and La-substitutions are the same, 
both $T_{coh}^{Yb}$ and $T_c^{Yb}$ remain much more robust with respect to disorder. This fact can be explained by noting that 
Yb atoms are in a mixed valence state and, therefore, must be correlated. The correlation may arise via local lattice deformations. 
To see how the impurity correlations may slow down the suppression of the coherence temperature, we may consider the characteristic length scale $R$ on which impurity distribution function significantly deviates from unity. 
The impurity distribution function determines the probability with which one can find one impurity at a certain distance from another. Within the Born approximation, one can show that there will be two contributions to the self-energy of the conduction electrons. One contribution, $\Sigma_{ii}$, corresponds to the scattering of electrons on the same impurity and, upon the averaging over disorder, this contribution is proportional to the concentration of impurities $n_{imp}$. The second contribution, $\Sigma_{ij}$, describes the scattering of electrons on two different impurities and, therefore, is proportional to $n_{imp}^2$. In the presence of impurity correlations, however, $\Sigma_{ij}$ becomes proportional to $n_{imp}^2R^3$. Thus, if the radius of correlations is large enough, $n_{imp}R^3\sim 1$, $\Sigma_{ij}$ becomes comparable with the first, linear in $n_{imp}$, self-energy correction $\Sigma_{ii}$. Consequently, within the large-$N$ mean field theory, one can show that impurity correlations may provide the "healing effect": the rate of suppression of the coherence temperature becomes strongly dependent on the impurity correlation length, $R$  \cite{Dzero2011}. 

\section{acknowledgements}
This work was supported by the National Science Foundation (grant NSF DMR-1006606 and DMR-0844115),  ICAM Branches Cost Sharing Fund from Institute for Complex Adaptive Matter, and Ohio Board of Regents (grant OBR-RIP-220573) at KSU, and by the U.S. Department of Energy (grant DE-FG02-04ER46105) at UCSD. MJ gratefully acknowledges financial support by the Alexander von Humboldt foundation.
\\
\\
$^{\star}$ These authors had equal contribution.

\end{document}